# Two-color ghost interference with photon pairs generated in hot atoms

Dong-Sheng Ding, Zhi-Yuan Zhou, Bao-Sen Shi*, Xu-Bo Zou and Guang-Can Guo

*Key Laboratory of Quantum Information, University of Science and Technology of China, Hefei 230026, China*

We report on an experimental observation of a two-photon ghost interference experiment. A distinguishing feature of our experiment is that the photons are generated via a non-degenerated spontaneous four-wave mixing process in a hot atomic ensemble; therefore the photon has narrow bandwidth. Besides, there is a large difference in frequency between two photons in a pair. Our works may be important to achieve more secure, large transmission capacity long-distance quantum communication.

Spontaneously parametric down-conversion (SPDC) in a nonlinear crystal [1, 2] as an efficient and widely used way of generating an entangled two-photon state, has been used in many interesting experiments [3-5]. Some of the most intriguing effects of a two-photon entangled state are quantum ghost interference and imaging [6, 7], the spatial non-classical correlation, such as transverse position and momentum, is used in these phenomena. Besides, such spatial correlation may have important applications in quantum information field, for example, it could be used to improve information transmission capacity and to achieve more secure key distribution [8] by establishing a high-dimensional entanglement. One disadvantage is that the photon generated by SPDC has so wide bandwidth (~THz) that it can't effectively couple with atoms (natural bandwidth: ~MHz) [9]. Such coupling is a prerequisite for building up a quantum repeater based on an atomic system [10]. Recently, another way for generating a photon pair, based on spontaneously Raman scattering (SRS) [11, 12] or spontaneous four-wave mixing (SFWM) [13-15] in an atomic ensemble, attracts peoples attentions due to the fact that the photon generated by this way has very narrow bandwidth. Although the non-classical temporal correlation between the photons generated through SRS or SFWM has been studied in many experiments [16-18], there is no any report about the spatial correlation between the photons. In this letter, we experimentally investigate the spatial correlation by performing a ghost interference experiment. We observe a clear two-slit interference pattern. The experimental result is well in agreement with our theoretical prediction. Our result shows that there is the spatial correlation between the photons in a pair. Besides, the correlated photons have a large difference in wavelength: one photon is in telecomband (1530 nm) and the other is at 780 nm, corresponding to the D2 of the rubidium (Rb) atom. Such a photon pair may be used to realize a quantum repeater for long-distance communication in a fiber system. Although a two-color image has been experimentally observed in Ref. 19 with a photon pair generated through SPDC, there is no any report about the two-color ghost interference using the photons generated in atoms. The combination of the spatial correlation between photons in a pair, suitable wavelength and the narrow bandwidth the photon has may help us realize a more secure, large transmission information capacity long-distance quantum communication.

The correlated photon pairs used in our experiment are generated through SFWM in $^{85}$Rb atoms with a ladder-type configuration shown in Fig. 1(a). The ladder-type configuration consists of one ground state |3> ($5S_{1/2}$), one intermediate state |2> ($5P_{3/2}$) and one upper state |1> ($4D_{5/2}$). The transition frequency between the ground state and the intermediate state corresponds to the D2 line of $^{85}$Rb, and the transition between the intermediate state and the upper state can be driven by a laser at 1529.4 nm. We use two beams named pump 1 (1529.4 nm) and pump 2 (780 nm) as two inputs to generate a non-degenerated non-classical correlated photon pair. The wavelengths of the two generated photons in a pair are 1529.4 nm and 780 nm respectively. $\Delta$ is the detuning between the frequencies of the pump 1 and the transition of $5P_{3/2} \rightarrow 4D_{5/2}$.

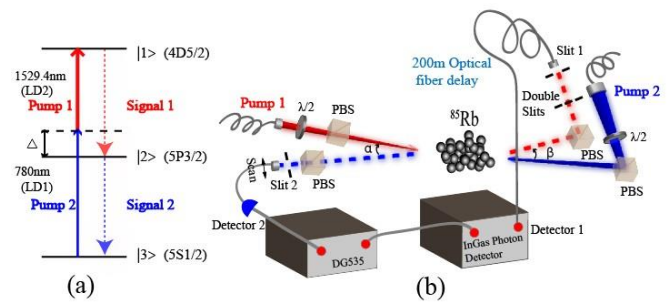

FIG. 1 (Color online) (a) Energy levels diagram of $^{85}$Rb used in our experiment. $\Delta$ is the detuning between the frequency of Pump 2 and the transition frequency between |3> and |2>, blue shift of $\Delta$ represents positive in our experiment. (b) Experiment setup. Signal 2 photons are collected by a multi-mode fiber, signal 1 photons are collected by a single-mode fiber. $\lambda/2$: half of wave plates; PBS: polarization beam splitter; DG535: delay generator; Detector 1: InGaAs Photon Detector; Detector 2: Avalanche diode.

The experimental setup is shown in Fig. 1(b). A 5-cm $^{85}$Rb vapour cell containing isotopically-pure $^{85}$Rb is used. A cw laser at 780 nm from an external-cavity diode laser (DL100, Toptica) is input to the Rb cell as pump 2 and a cw laser beam at 1529.4 nm from another external-cavity diode laser (DL100, Prodesign, Toptica) is used as pump 1. The powers of pumps 1 and 2 can be adjusted through a half wave plates and a polarization beam splitter respectively. The polarizations of pump 1 and pump 2 are orthogonal. The angle between two pump lights is about 1.27°. Both pumps are focused and their waist widths at the center of Rb cell are about 0.6 mm and 0.35 mm respectively. There is a small angle α(β) between pump 1 (pump 2) and the generated photon at 780 nm (1529.4 nm), and α≈β=2.26°. Two lenses (not shown in Fig. 1(b)) act as a telescope to collect all generated photons at 780 nm (signal 2) into a single-mode fiber. The fiber is connected to the detector 2 (PerkinElmer SPCM-AQR-15-FC). Note that the lenses merely serve the purpose of collecting photons. The generated photon at wavelength 1529.4 nm (signal 1) is collected by a multi-mode fiber. After about 1000 ns time delay caused in a 200-m long fiber, the signal 1 is sent to the detector 1 working at gated mode (InGaAs Photon detector with 8% detection efficiency). Signals 1 and 2 have orthogonal polarization directions. A two-slit mask with the slit width of 0.2 mm and the distance d between two slits of 0.5 mm is inserted in the signal 1 beam. The slit 1 with width of 1 mm in the signal 1 serves as a bucket detector, and the slit 2 with width of 0.2 mm in the signal 2 beam served as a point detector. The detector 1 is triggered by the delayed electronic pulse from the detector 2. Through adjusting the delay time by a delay generator (DG535), we could obtain the cross-correlation function with high accuracy. Firstly, we check the temporal correlation between the photons. In this case, the double-slit mask is removed. If the photons in a pair are in non-classical correlated in time domain, the Cauchy-Schwarz inequality should be violated. Our experimental results [20] show that the Cauchy-Schwarz inequality is strongly violated by the factor of R=48±12, where, $R=g_{s1s2}(t)^2/g_{s1s1}g_{s2s2}$, $g_{s1s2}(t)$, $g_{s1s1}$ and $g_{s2s2}$ are cross-correlation and auto-correlations of the photons respectively. R> 1 means the non-classical correlation existed between two photons. Our result clearly shows that there is the non-classical correlation in time domain between photons in a pair. The estimated bandwidth of the photon is about 1.5 ns according to the cross-correlation measurement.

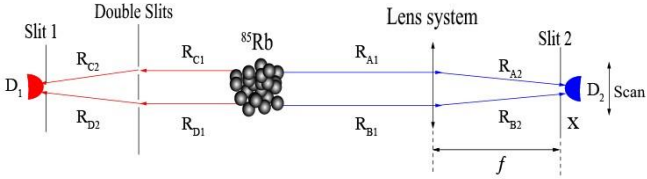

FIG. 2 (Color online) Our simple theoretical diagram for fitting experimental data. $f$ is focal length of lens system; In our experiment, $f$ is about 220 mm.

In order to understand the physics clearly of the experiment, we unfold the setup shown in Fig. 1(b), and show it in Fig. 2. We consider such a physical process in the experiment: when signal 1 photon passes through the two-silt mask while the signal 2 gets to $D_2$. We could treat the atomic ensemble as a "geometrical reflection" mirror, as pointed out in Ref. 6. According to the ghost interference theory shown in Ref. 6, and if we assume the optical paths in upper and lower paths $R_{C1}=R_{D1}$, $R_{C2}=R_{D2}$ and $R_{A1}=R_{B1}$, then the difference of the optical path $R_{A1}+R_{A2}+R_{C1}+R_{C2}$ and $R_{B1}+R_{B2}+R_{D1}+R_{D2}$ equals to the difference of $R_{A2}$ and $R_{B2}$, and the cross-coincidence count is

$$R_c \propto \text{sinc}^2\left(\frac{x\pi a}{\lambda_2 f}\right)\cos^2\left(\frac{x\pi d}{\lambda_2 f}\right) . \quad (1)$$

Where, $\lambda_2$ is the wavelength of signal 2, α is the slit width and $d$ is the distance between two slits.

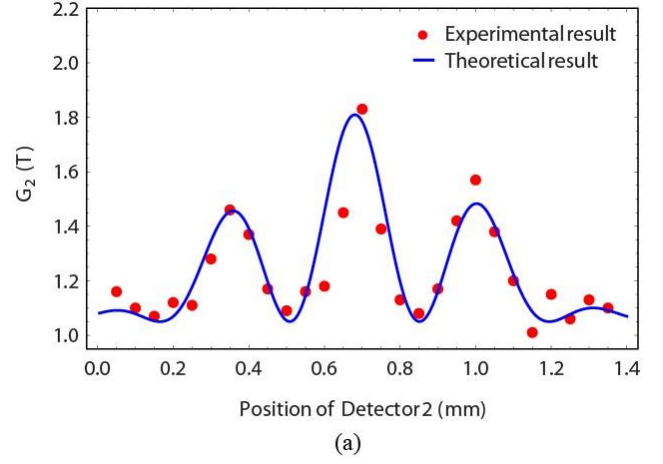

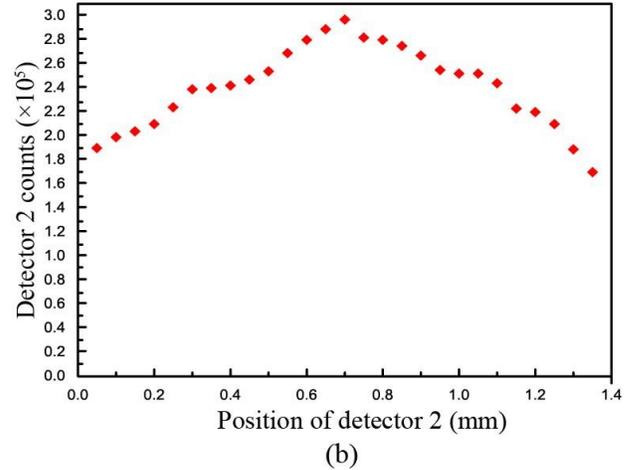

FIG. 3 (Color online) (a) Double-slit interference pattern with two-color photon through scanning the position of Detector 2. Red points are experimental data, and blue line is theoretical fit using Eq. (1). (b) Single counts of signal 2 photons with scanning of Detector 2.

In our experiment, Δ=2.5 GHz, the temperature of atoms is 110 °, the powers of pump 1 and pump 2 are 28 mW and 50 μW respectively. We set the two-photon detuning within the range of Doppler broadening by adjusting the detuning of pump 1. The signal 1 photons are coupled into a single-mode fiber with 50% coupling efficiency and the signal 2 photons are collected by a multi-mode fiber with 90% coupling efficiency. We scan detector 2 with each step of 50 μm in transverse direction to the propagation direction of the signal 1 to measure the cross-correlation coincidence count between the detectors 1 and 2, and monitor the single count of detector 2 simultaneously. The recorded data are shown in Fig. 3(a) and Fig. 3(b) respectively. It is very clear that coincidence counts show a well-known double-slit interference fringe, at the same time, single counts have no obvious oscillation. In Fig. 3(a), the dots are experimental data, and the solid line is fitted curve by Eq. (1) using our experimental parameters: $\lambda_2$=780 nm, α=200 μm, d=500 μm and $f$ =220 mm. The theoretical fit is in agreement very well with our experimental data. The relatively low interference visibility is mainly due to the low spatial correlation between the photons. It is well known if the divergence of the generated photon is much larger than $\lambda_2$/d, then a high interference visibility could be obtained. The divergence of the photon in our experiment is about 3.2 mrad according to the experimental conditions, almost equal to $\lambda_2$/d=3 *mrad*, therefore the visibility obtained experimentally is not high. In principle, we could improve the visibility by increasing the size of the pump beam, but it will reduce the photon production rate when the pump power is fixed. It also increases the difficulty to how to completely reduce the noise from the pump laser, inducing the low signal-to-noise ratio. Besides, the size of the detector also influences the visibility. A small detector size will improve the visibility.

The ghost interference observed in this experiment utilizes the spatial correlation of the biphoton, but we have to say that whether the phenomenon is caused by non-classical spatial correlation has not been demonstrated yet. It needs the further proof, like done in Refs. 21, 22, will be our next work. The two-photon ghost interference can be observed with a classical source, and is demonstrated in Ref. 23. Ghost imaging has also been observed by classical lights [24-26].

In conclusion, we perform a two-color ghost interference experiment with a photon pair generated through a SFWM in an atomic system and observe a clear two-slit interference pattern. The result is well agreement with our theoretical prediction. The big difference between the well-known works and ours is that the photon pair used in our experiment is generated via a SFWM in a hot atomic vapor. Our result may useful for achieving more secure, large transmission capacity long-distance quantum communication.

**Acknowledgements**

We thank Dr. Wei Chen and Bi-Heng Liu for technique support. This work was supported by the NSFC (Grant Nos. 10874171, 11174271), the National Fundamental Research Program of China (Grant No. 2011CB00200).

* *drshi@ustc.edu.cn*

*References*
1. P. Kwiat, K. Mattle, H. Weinfurter, and A. Zeilinger, Phys. Rev. Lett. **75**, 4337-4340 (1995).
2. F. Y. Wang, B. S. Shi and G. C. Guo, Opt. Lett. **33**, 2191-2193 (2008).
3. C. K. Hong, Z. Y. Ou and L. Mandel, Phys. Rev. Lett. **59**, 2044-2047 (1987).
4. C. H. Bennett, C. Crépeau, R. Jozsa, A. Peres, and William K. Wootters, Phys. Rev. Lett. **70**, 1895-1898 (1993).
5. D. Bouwmeester, J.-W. Pan, K. Mattle, M. Eibl, H. Weinfurter, A. zeilinger, Nature, **390**, 575 (1997).
6. D. V. Strekalov, A. V. Sergienko, D. N. Klyshko, and Y. H. Shih, Phys. Rev. Lett. **74**, 3600 (1995).
7. T. B. Pittman, Y. H. Shih, D. V. Strekalov, and A. V. Sergienko, Phys. Rev. A. **52**, R3429 (1995).
8. H. Bechmann-Pasquinucci, and W. Tittel, Phys. Rev. A. **61**, 062308 (2000).
9. B. S. Shi, F. Y. Wang, C. Zhai, and G. C. Guo, Opt. Commun. **281** 3390 (2008).
10. L.-M. Duan, M. D. Lukin, J. I. Cirac and P. Zoller, Nature, **414**, 413, (2001).
11. A. Kuzmich, W. P. Bowen, A. D. Boozer, A. Boca, C. W. Chou, H. J. Kimble, Nature, **423**, 731 (2003)
12. S. Chen, Y. O Chen, T. Strassel, Z. Y. Yuan, B. Zhao, J. Schmiedmayer, and J. W. Pan, Phys. Rev. Lett. **97**, 173004 (2006).
13. S. Du, P. Kolchin, C. Belthangady, G. Y. Yin, and S. E. Harris, Phys. Rev. Lett. **100**, 183603 (2008).
14. Q. F. Chen, B. S. Shi, M. Feng, Y. S. Zhang, G. C. Guo, Opt. Express. **16**, 21708 (2008).
15. X. S. Lu, Q. F. Chen, B. S. Shi and G. C. Guo, Chin. Phys. Lett. **26**, 064204 (2009).
16. D. N. Matsukevich and A. Kuzmich, Science, **306**, 663 (2004).
17. A. G. Radnaev, Y. O. Dudin, R. Zhao, H. H. Jen, S. D. Jenkins, A. Kuzmich, and T. B. A. Kennedy. Nature Physics, **6**, 894-899, (2010).
18. K. S. Choi, H. Deng, J. Laurat and H. J. Kimble, Nature, **452**, 67 (2008).
19. S. Karmakar and Y. Shih, Phys. Rev. A. **81**, 033845 (2010).
20. D. S. Ding, Z. Y. Zhou, B. S. Shi, X. B. Zou, and G. C. Guo, Opt. Express. **20**, 11433-11444 (2012).
21. J. C. Howell, R. S. Bennink, S. J. Bentley, and R. W. Boyd, Phys. Rev. Lett. **92**, 210403 (2004).
22. M. D'Angelo, Y. H. Kim, S. P. Kulik, and S. Shih, Phys. Rev. Lett. **92**, 233601 (2004).
23. F. Ferri, D. Magatti, A. Gatti, M. Bache, E. Brambilla, and L. A. Lugiato, Phys. Rev. Lett. **94**, 183602 (2005).
24. R. S. Bennink, S. J. Bentley, and R. W. Boyd, Phys. Rev. Lett. **89**, 113601 (2002).
25. X. H. Chen, Q. Liu, K. H. Luo, L. A. Wu, Opt. Lett. **34**, 695 (2009).
26. A. Gatti, E. Brambilla, M. Bache and L. A. Lugiato, Phys. Rev. Lett. **93**, 093602 (2002).